\documentclass[aps,prd,preprint,tightenlines,superscriptaddress,nofootinbib]{revtex4}
\usepackage{amsmath,amssymb,url}
\usepackage{graphicx,tabularx}
\usepackage{array,dsfont}

\def\bq{\begin{equation}}
\def\eq{\end{equation}}
\def\ba{\begin{eqnarray}}
\def\ea{\end{eqnarray}}
\newcommand{\sla}[1]{/\!\!\!#1}

\newcommand{\LL}{\mathcal{L}}

\newcommand{\Tr}{\mathop{\rm Tr}}

\newcommand{\hc}{\text{h.c.}}

\renewcommand{\Im}{\text{Im}}

\newcommand{\cx}{c_x}

\newcommand{\TeV}{{\ensuremath\rm TeV}}

\newcommand{\fb}{{\ensuremath\rm fb}}

\begin{document}

\date{\today}

\preprint{DESY 06-055}

\title{Distinguishing Little-Higgs Product and Simple Group models\\
at the LHC and ILC}

\author{W.~Kilian}
\email{wolfgang.kilian@desy.de}
\affiliation{%
 Fachbereich Physik, University of Siegen, D--57068 Siegen, Germany}
\affiliation{Deutsches Elektronen-Synchrotron DESY, D--22603 Hamburg,
Germany}
\author{D.~Rainwater}
\email{rain@pas.rochester.edu}
\affiliation{Department of Physics and Astronomy, University of 
Rochester, Rochester, NY 14627, USA}
\author{J.~Reuter}
\email{juergen.reuter@desy.de}
\affiliation{Deutsches Elektronen-Synchrotron DESY, D--22603 Hamburg,
Germany}

\begin{abstract}
We propose a means to discriminate between the two basic variants of
Little Higgs models, the Product Group and Simple Group models, at the
next generation of colliders.  It relies on a special coupling of
light pseudoscalar particles present in Little Higgs models, the
pseudo-axions, to the $Z$ and the Higgs boson, which is present only
in Simple Group models.  We discuss the collider phenomenology of the
pseudo-axion in the presence of such a coupling at the LHC, where
resonant production and decay of either the Higgs or the pseudo-axion
induced by that coupling can be observed for much of parameter space.
The full allowed range of parameters, including regions where the
observability is limited at the LHC, is covered by a future ILC, where
double scalar production would be a golden channel to look for.
\end{abstract}

\maketitle


\section{Introduction}

Little Higgs models~\cite{LH-orig} have been proposed as a solution to
the hierarchy problem, the vast difference between the scales of
electroweak symmetry breaking and the scale where quantum effects of
gravity can no longer be neglected.  The Higgs boson emerges as a
pseudo-Goldstone boson of spontaneously broken chiral symmetries of a
strongly interacting theory in the multi-TeV regime.  The
quadratically divergent contributions to the Higgs mass are canceled
-- unlike in supersymmetry -- by particles of like statistics, chiral
multiplet partners of the Standard Model (SM) fermions.  The
collective breaking of these symmetries generates a
quadratically-divergent correction to the Higgs mass only at two-loop
order, such that the models remain weakly interacting at the TeV
scale, thereby satisfying electroweak precision data constraints.

Ideas about possible UV completions
exist~\cite{UV-strong,UV-iterated,UV-SUSY}, but we ignore them in this
paper and take Little Higgs models as low-energy effective theories.
Various realizations of the Little Higgs symmetry structure have been
proposed~\cite{LH-min,minmoose,simple,LH-mod,simplest}, grouped quite
generally into two classes~(cf.~e.g.~\cite{smoking,CPNSH}).  The
original models are mostly Product Group Models, where the electroweak
gauge group is extended to a product gauge group that contains $SU(2)$
subgroups in at least two distinct factors.  The physical $W$ and $Z$
bosons are mixtures of the gauge bosons from both factors.  The global
symmetry group is a simple group that includes the full gauge
structure [e.g., $SU(5)\supset SU(2)\times SU(2)\times U(1)$], and the
Higgs is part of an irreducible multiplet of pseudo-Goldstone bosons.
In Simple Group Models, the electroweak $SU(2)$ gauge group is
embedded in a simple (e.g., $SU(N)$) gauge group, while the global
symmetry group has a product structure, and the Higgs is a mixture of
the components of several (at least two) independent irreducible
representations of the factor groups.

Although these models remain weakly interacting at the TeV scale,
nevertheless electroweak precision observables pose constraints on
their parameter space~\cite{Csaki,Hewett,KR03,CDO03}~\footnote{There
are significant corrections to mass terms and couplings in the
Littlest Higgs model~\cite{Buras:2006wk,Boersma:2006hw}.}.  The decay
constant $F$ of the Pseudo-Goldstone bosons which sets the mass scale
for the additional vector bosons, scalars and fermions in Little Higgs
models, has to lie in the range $\sim 1-4$~TeV, where the upper bound
comes from naturalness considerations.  Usually the masses of the
additional particles lie in the range of several hundred GeV up to a
few TeV, and are therefore (partly) directly accessible at the
LHC~(for an overview about Little Higgs phenomenology,
cf. Ref.~\cite{LH_summ}).  The proof of the Little Higgs mechanism
within these models relies on the discovery of the nonlinear Goldstone
boson structure of the Higgs and the cancellation mechanism of the
quadratic divergences based on measurements of the couplings.  This
will not be possible in all cases at the LHC, but should be at a
future international $e^+e^-$ collider (ILC).  An ILC would in any
case be necessary for precision measurements.  To reveal a special
group theoretical realization of Little Higgs models is also quite
difficult and relies on precise branching ratio measurements for the
Higgs, the extra gauge bosons and fermionic states.  A method to
distinguish between Simple and Product Group Models, maybe even
already at the LHC, would be highly welcome.

We propose a method based on the discovery and the properties of light
pseudoscalar states in the spectrum, the Little Higgs
pseudo-axions~\cite{pseudoaxion}.  We first recall their properties,
then show that a specific pseudo-axion coupling, to the Higgs and the
$Z$ bosons, can occur only in Simple Group Models.  It could thus be
used as a discriminator between the two Little Higgs model classes.
We then discuss relevant phenomenology at the LHC and ILC.


\section{A discriminator for Simple and Product Group Models}
\label{sec:disc}

In Little Higgs models, there can occur spontaneously broken
(approximate) global $U(1)$ symmetries, corresponding to diagonal
generators of the overall non-abelian global symmetry
group~\cite{pseudoaxion}.  To each of these $U(1)$ factors, which
might well be anomalous, corresponds a (pseudo-) Goldstone boson which
couples to fermions like a pseudoscalar, analogous to the
$\eta^{(\prime)}$ meson in chiral symmetry breaking of QCD.  Such a
particle is an electroweak singlet, which gets a mass through explicit
symmetry breaking terms and the Coleman-Weinberg potential.  All
couplings to SM particles are therefore suppressed by the ratio of the
electroweak and the Little Higgs scale, $v/F$.  All couplings to SM
gauge bosons are induced by anomalous triangle loops.  As a typical
example for these particles, we show the situation for the so-called
Simplest Little Higgs or $\mu$--model~\cite{simplest} in
Fig.~\ref{fig:par_space}.  Since this is a Simple Group Model, there
are two multiplets of Goldstone bosons connected by a mixing angle
$\tan\beta$ as in two-Higgs-doublet models.  Furthermore, there is an
explicit breaking of the global Little Higgs symmetries, analogous to
the $\mu$ term in the MSSM.  Pseudo-Axion phenomenology at the LHC and
future ILC and photon colliders was discussed in~\cite{pseudoaxion}.
There it was shown that in the Simplest Little Higgs there is a
tree-level coupling $Z$--$H$--$\eta$ of the pseudo-axion to the Higgs
and the $Z$, which can only arise by electroweak symmetry breaking,
and is enhanced by $\tan\beta$.  On the other hand, it was shown that
this coupling is absent in the simplest candidate of the Product Group
Models, the Littlest Higgs.

We now show that the existence of such a $Z$--$H$--$\eta$ coupling is
a property of Simple Group models, and that it cannot appear in
Product Group models.  Hence, it serves as a discriminator between the
two categories.  The crucial observation is that the
matrix-representation embedding of the two non-Abelian $SU(2)$ gauge
groups, and especially of the two $U(1)$ factors within the
irreducible multiplet of the pseudo-Goldstone bosons of one simple
group (e.g. $SU(5)$ in the Littlest Higgs), is responsible for the
non-existence of this coupling in Product Group models.  It is exactly
the mechanism which cancels the quadratic one-loop divergences
between the electroweak and heavy $SU(2)$ gauge bosons which cancels
this coupling.  In Simple Group models the Higgs mass term
cancellation is taken over by enlarging $SU(2)$ to $SU(N)$, and the
enlarged non-Abelian rank structure cancels the quadratic divergences
in the gauge sector -- but no longer forbids the $Z$--$H$--$\eta$
coupling.

To leading order in the pseudo-axion field $\eta$, the
parameterization of the Goldstone boson manifold in the effective
Lagrangian does not depend on the basis choice of the broken
generators~\cite{Callan:1969sn}.  Therefore one can take it
proportional to the unit matrix, i.e. factor it out from the matrix of
pseudo-Goldstone boson.  This corresponds to a separation of the
special $U(1)_\eta$ group (cf. also the discussion in
Refs.~\cite{simplest} and~\cite{pseudoaxion}).  Here we use
$\xi=\exp\left[i\eta/F\right]$ for the pseudo-axion field and
$\Sigma=\exp\left[i\Pi/F\right]$ for the non-linear representation of
the remaining Goldstone multiplet $\Pi$ of Higgs and other heavy
scalars.  Then, for Product Group Models, the kinetic term may be
expanded as
\begin{align}
  \label{LL-kin}
  \LL_{\text{kin.}} \sim F^2
  \Tr \left[ (D^\mu (\xi \Sigma)^\dagger (D_\mu (\xi \Sigma)) \right]
  &= \ldots + F^2 (\partial_\mu \xi) \xi^\dagger
  \Tr \left[ (D^\mu \Sigma)^\dagger \Sigma \right]
  + \hc 
\nonumber\\
  &= \ldots - 2F(\partial_\mu \eta)\,
  \Im \Tr \left[  (D^\mu \Sigma)^\dagger \Sigma \right]
  + O(\eta^2),
\end{align}
where we write only the term with one derivative acting on $\xi$ and
one derivative acting on $\Sigma$.  This term, if nonzero, is the only
one that can yield a $Z$--$H$--$\eta$ coupling.

We now use the special structure of the covariant derivatives in
Product Group Models, which is the key to the Little Higgs mechanism:
\bq
  D_\mu \Sigma = \partial_\mu \Sigma + A_{1,\mu}^a \left( T_1^a \Sigma
  + \Sigma (T_1^a)^T \right) + A_{2,\mu}^a \left( T_2^a \Sigma
  + \Sigma (T_2^a)^T \right), 
\eq
where $T_i^a, i=1,2$ are the generators of the two independent $SU(2)$
groups (extra gauge structure does not matter), and $A_{i,\mu}^a =
W^a_\mu$ + heavy fields in a suitable normalization
(cf. Ref.~\cite{KR03}).  Neglecting the heavy gauge fields and
extracting the electroweak gauge bosons, we have
\begin{align}
\Tr\left[  (D^\mu \Sigma)^\dagger \Sigma \right] 
&\sim W_\mu^a \Tr \left[ (\Sigma^\dagger (T_1^a + T_2^a)\Sigma
                        +(T_1^a + T_2^a)^* \right]
\notag \\ 
&\quad = W_\mu^a \Tr \left[ (T_1^a + T_2^a) + (T_1^a + T_2^a)^* \right]
= 0.
\end{align}
This vanishes due to the zero trace of $SU(2)$ generators.  The same
is true when we include additional $U(1)$ gauge group generators such
as hypercharge, since their embedding in the global simple group
forces them to be traceless as well.  We conclude that the coefficient
of the $Z$--$H$--$\eta$ coupling vanishes to all orders in the $1/F$
expansion.

Next, we consider the kinetic term for Simple Group Models, where we
use the following notation for the nonlinear sigma fields:
$\Phi\zeta$, where $\Phi=\exp[i\Sigma/F]$ and $\zeta=\left(0,\ldots
0,F\right)^T$ is the vacuum expectation vector directing in the $N$
direction for an $SU(N)$ simple gauge group extension of the weak
group.  Thus, in Simple Group Models the result is the $N,N$ component
of a matrix:
\begin{align}\label{kin:simplegroup}
\LL_{\text{kin.}} \sim F^2 D^\mu (\zeta^\dagger \Phi^\dagger) D_\mu
 (\Phi\zeta) =&\; \ldots + 
 \frac{i}{F} (\partial_\mu \eta) \zeta^\dagger
\left( \Phi^\dagger (D_\mu \Phi) - (D_\mu \Phi^\dagger) \Phi \right)
\zeta \notag \\ =&\; \ldots +  i F (\partial_\mu \eta) \left(
 \Phi^\dagger (D_\mu \Phi) - (D_\mu \Phi^\dagger) \Phi \right)_{N,N} \; .
\end{align}
To further evaluate this term, we separate the last row and column in
the matrix representations of the Goldstone fields $\Sigma$ and gauge
boson fields $\mathds{V}_\mu$:
\bq\label{matrixdecomp}
\Sigma = 
\begin{pmatrix}
  0 & h \\ 
  h^\dagger & 0 
\end{pmatrix}, \qquad \qquad
\mathds{V}_\mu =
\begin{pmatrix}
  \mathds{W}_\mu & 0 \\ 
  0 & 0 
\end{pmatrix} + \text{heavy vector fields}
\eq
The Higgs boson in Simple Group Models sits in the off-diagonal
entries (one doublet for the Simplest LH and a pair of doublets for
the Original Simple Group model), while the electroweak gauge bosons
reside in the upper left corner.

With the Baker-Campbell-Hausdorff identity, one gets for the term in
parentheses in Eq.~(\ref{kin:simplegroup}):
\begin{multline}\label{eq:vexpansion}
  \mathds{V}_\mu + \frac{i}{F} \lbrack \Sigma, \mathds{V}_\mu \rbrack 
  - \frac{1}{2F^2} \lbrack \Sigma , \lbrack \Sigma, \mathds{V}_\mu
  \rbrack \rbrack + \ldots \\
  =\begin{pmatrix}
      \mathds{W}_\mu & 0 \\ 
      0 & 0 
   \end{pmatrix}
   + \frac{i}{F} 
  \begin{pmatrix}
   0 & - \mathds{W}_\mu h \\ h^\dagger \mathds{W}_\mu & 0 
  \end{pmatrix} - \frac{1}{2F^2} 
  \begin{pmatrix}
    h h^\dagger \mathds{W} + \mathds{W} h h^\dagger & 0
    \\ 0 & - 2 h^\dagger \mathds{W} h 
  \end{pmatrix} + \ldots
\end{multline}
The $N,N$ entry can only be nonzero from the third term on.  This can
be understood as follows.  In the first term, the $N,N$ component
cancels by the help of the multiple Goldstone multiplets present in
the Simple Group Models; it would be a mixing between the $\eta$ and
the Goldstone boson(s) for the $Z'$ state(s).  If the $N,N$ component
of the second term were nonzero, it would induce a $Z$--$H$--$\eta$
coupling without insertion of a factor $v$.  This is forbidden by
electroweak symmetry.  To see this, it is important to note that
$W^3\equiv A^3$ in the Simple Group Models while the hypercharge boson
$B$ is a mixture of the diagonal $U(1)$ generator and the left-over
diagonal generators in $SU(N)$ ($T_8$ in $SU(3)$, $T_{12}, T_{15}$ in
$SU(4)$, etc.).  The embedding of the Standard Model gauge group
always works in such a way that hypercharge is a linear combination of
the $T_{N^2-1}$ and $U(1)$ generators.  This has the effect of
canceling the $\gamma$ and $Z$ from the diagonal elements beyond the
first two positions, and preventing the diagonal part of
$\mathds{W}_\mu$ from being proportional to $\tau^3$.  By using
Eq.~(\ref{matrixdecomp}) one easily sees that
$\lbrack\Sigma,\mathds{V}_\mu\rbrack_{N,N}$ is zero.

The third term in the expansion yields a contribution to the
$Z$--$H$--$\eta$ coupling,
\bq
(\partial^\mu\eta)h^\dagger \mathds{W}_\mu h 
\sim v H Z_\mu\partial^\mu\eta \; .
\eq

Since in Simple Group models the nonlinear Goldstone boson multiplets
always come in pairs, there may still be a cancellation between
parameters that makes the prefactor vanish; however, this occurs only
for degenerate parameter sets.  By contrast, in Product Group models,
the absence of this coupling is a property of the symmetry structure
and occurs for all possible parameter values.

As a concrete example, we derive the $Z$--$H$--$\eta$ coupling in the
Simplest Little Higgs or $\mu$--model~\cite{simplest}.  In this model,
the EW group is enlarged to a gauged $SU(3)\times U(1)$.  There are
two nonlinear sigma fields, each of which parameterizes a coset space
$U(3)/U(2)$:
\bq
\Phi_1 = \exp\left[  i \tan\beta \, \Theta\right]
\begin{pmatrix} 0 \\ 0 \\ F_1 \end{pmatrix}
\qquad
\Phi_2 = \exp\left[ -i \cot\beta \, \Theta\right]
\begin{pmatrix} 0 \\ 0 \\ F_2 \end{pmatrix}
\eq
where
\bq
\Theta = \frac{1}{F} \left\{ \frac{\eta}{\sqrt{2}} +
                             \begin{pmatrix}
                               \begin{array}{cc}
                                 0 & 0 \\ 0 & 0
                               \end{array} & h^* \\
                               h^T & 0
                             \end{pmatrix} \right\} \, ,
\qquad
F^2 = F_1^2 + F_2^2 \, ,
\qquad {\rm and} \;\;\;
\tan\beta = \frac{F_2}{F_1}\;.
\eq
The gauge boson multiplet $\mathds{V}_\mu$ consists of the $SU(3)_w$
gauge bosons and the $U(1)$ gauge boson $B_x$.  It decomposes into the
electroweak gauge bosons $W^\pm$, $Z$, the photon $A$, and heavy
vector bosons $X^\pm,X^0,Y^0,Z^{\prime 0}$.  In the covariant
derivative acting on the Higgs multiplets, they enter via the
matrix~\cite{simplest}
\begin{multline}
  \frac12 g A^a\lambda^a - \frac13 g_x B_x = \\ 
  \begin{pmatrix}
    e A + \frac{g}{2c_w} (2 - 3 s_w^2) Z + \frac{g}{2 \sqrt3 \cx} (1 -
       3 s_x^2) Z' 
       & \sqrt2 W^- & i\sqrt2 X^- \\
    \sqrt2 W^+ & - \frac{g}{2c_w} Z + \frac{g}{2 \sqrt3 \cx} (1 -
       3 s_x^2) Z' 
       & Y^0 + iX^0    \\
    -i\sqrt2 X^+ 
       & Y^0 - iX^0 & - \frac{g}{\sqrt3 c_x} Z'
  \end{pmatrix},
\end{multline}
%
which results in the following couplings of the pseudo-axion $\eta$:
\bq
\LL_{VH\eta} = (\eta \, \overset{\leftrightarrow}{\partial^\mu} H)
\left[ \frac{m_Z}{\sqrt{2} F} N_2 (Z_\mu + c_w c_x Z'_\mu) +
  \frac{g}{2} N_2 c_\beta s_\beta X^0_\mu \right].
\eq
$N_2$ is defined as in Ref.~\cite{pseudoaxion}:
\bq
N_2 = \frac{F_2^2 - F_1^2}{F_1F_2} = \tan\beta - \cot\beta
\eq
As long as $F_1\neq F_2$, or $\tan\beta\neq 1$, there is a
$Z$--$H$--$\eta$ coupling as anticipated.

To calculate the $\eta$ production rates via gluon-gluon fusion at
hadron colliders, and the $\eta$ branching ratio to observable final
states, we also need its fermion couplings.  In the $\mu$ model, these
are as follows.  The couplings to top quark and heavy top partner
fermion $T$ are 
$g_{\eta tt}=\frac{m_tN_2}{\sqrt{2}F}-\frac{m_t^2N_1}{vM_T}$ and
$g_{\eta TT}=\frac{N_1m_t}{v}$, where
$N_1=\frac{F_1F_2}{F^2}\frac{\lambda_1^2-\lambda^2_2}{\lambda_1\lambda_2}$
and $\lambda_1$,$\lambda_2$ are the top Yukawa couplings of the model,
chosen to produce the observed top quark mass and minimize $m_T$ (cf.
Ref.~\cite{pseudoaxion}).  The $\eta b\bar{b}$ coupling is 
$g_{\eta bb}=-N_2m_b/\sqrt{2}F$, and has two interesting properties: 
it appears only at higher orders in the expansion of $\Phi_{1,2}$ (yet
remains ${\cal O}(v/F)$), and it is driven rapidly to zero the closer
$F_1$ and $F_2$ are (as $\tan\beta\to 1$).


\section{Phenomenology at the LHC}
\label{sec:LHC}

To discuss the possible LHC $Z$--$H$--$\eta$ coupling phenomenology,
we study the Simplest Little Higgs and adopt the parameter set of
Ref.~\cite{simplest}, consistent with existing EW and flavor data and
the preference for a light Higgs boson.  The relevant parameters are
$F_1$, $F_2$, $\mu$ and $\Lambda$.  For a generic scale of
$\Lambda=5$~TeV at which the Little Higgs effective theory breaks
down, our only free parameters are then $F_{1,2}$ and $\mu$, with the
constraint that $F_1$ not be as small as the EW scale, $F\gtrsim
2$~TeV from EW precision constraints (primarily $\Delta T$ and
four-fermion operators), and $F_2 \gtrsim F_1$ ($\tan\beta\gtrsim 1$)
to avoid too much mixing which would lead to fermion non-universality.
$F_1=F_2$ ($\tan\beta=1$) lies right at the edge of the limits on the
latter, but this point zeroes the $Z$--$H$--$\eta$ coupling and is of
trivial interest.  We vary $\mu$ over a slightly broader range than
the calculated $m_H$ would suggest is allowed or favored by data.
\begin{figure}[hb!]
\begin{center}
\begin{tabular}{lr}
\includegraphics[scale=0.8]{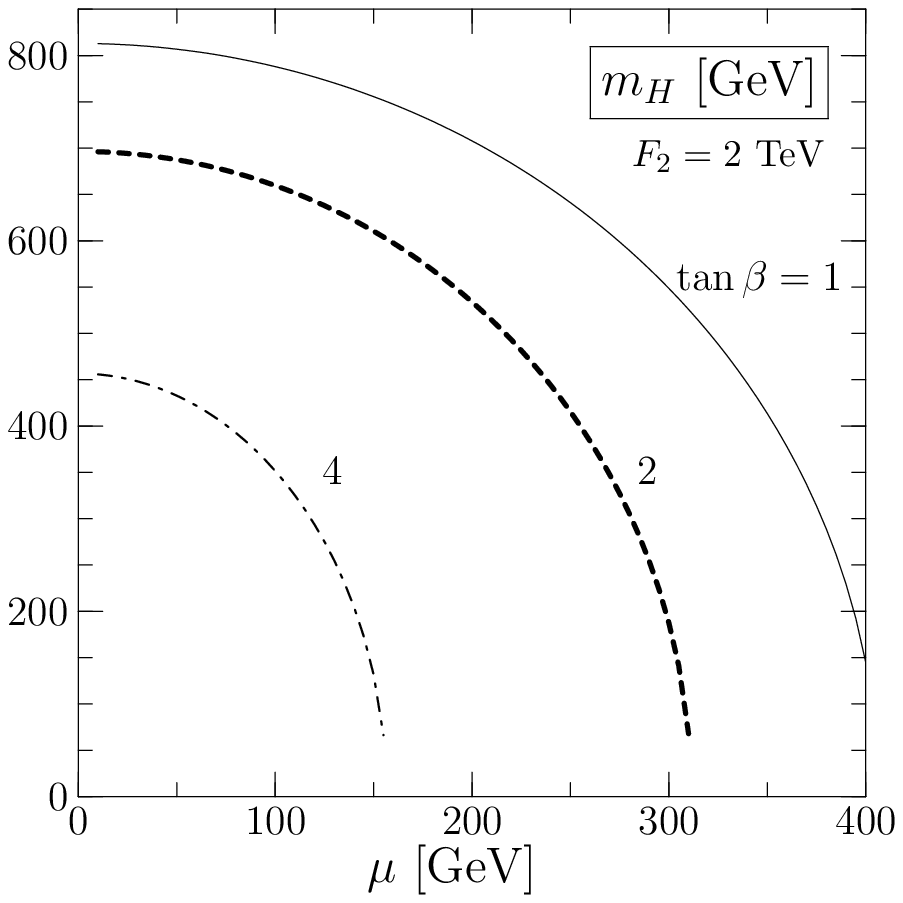}  \qquad &
\includegraphics[scale=0.8]{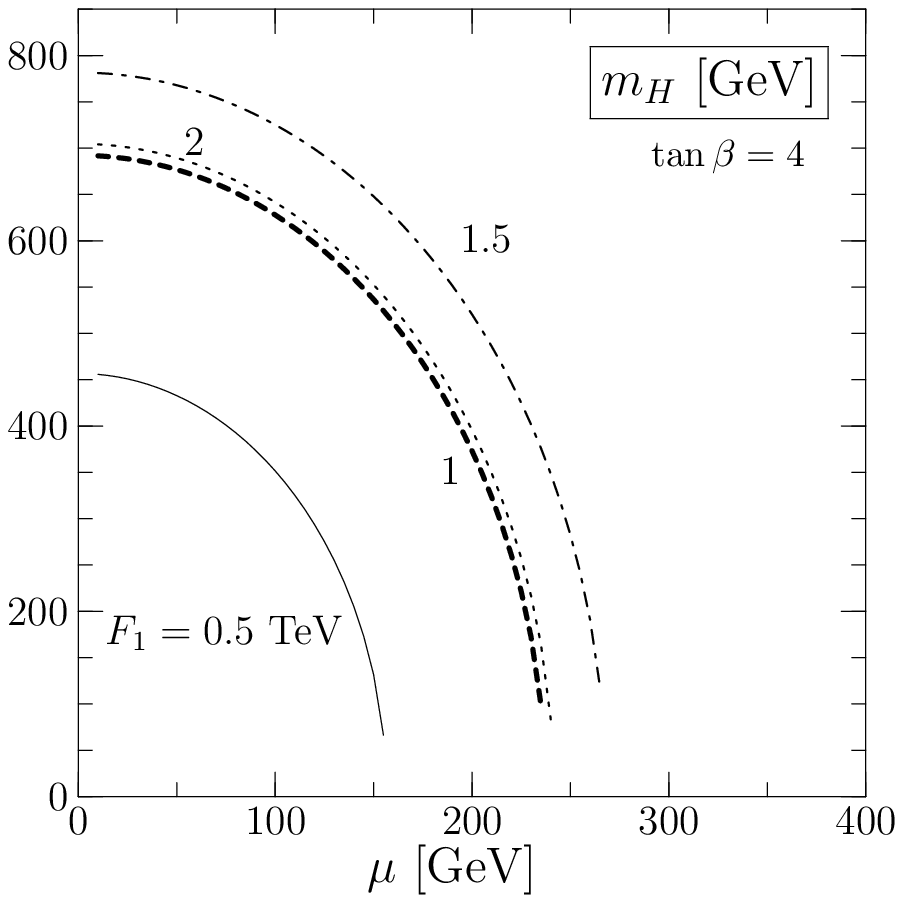}  \\[2mm]
\includegraphics[scale=0.8]{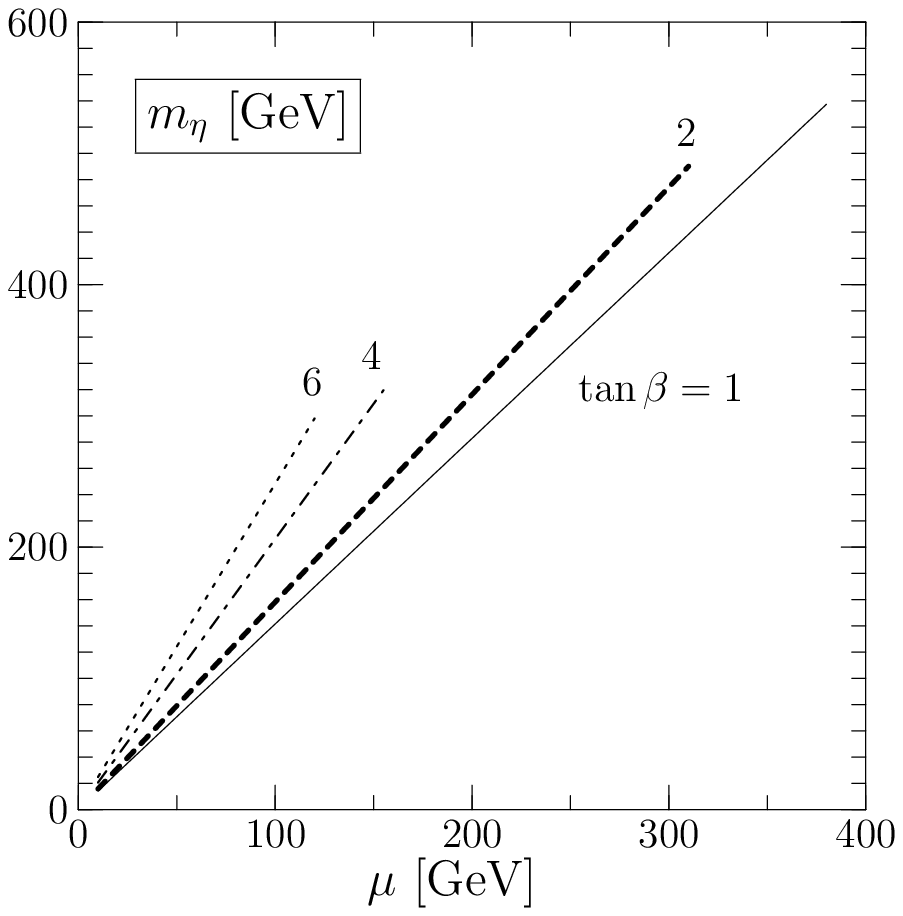} \qquad &
\includegraphics[scale=0.8]{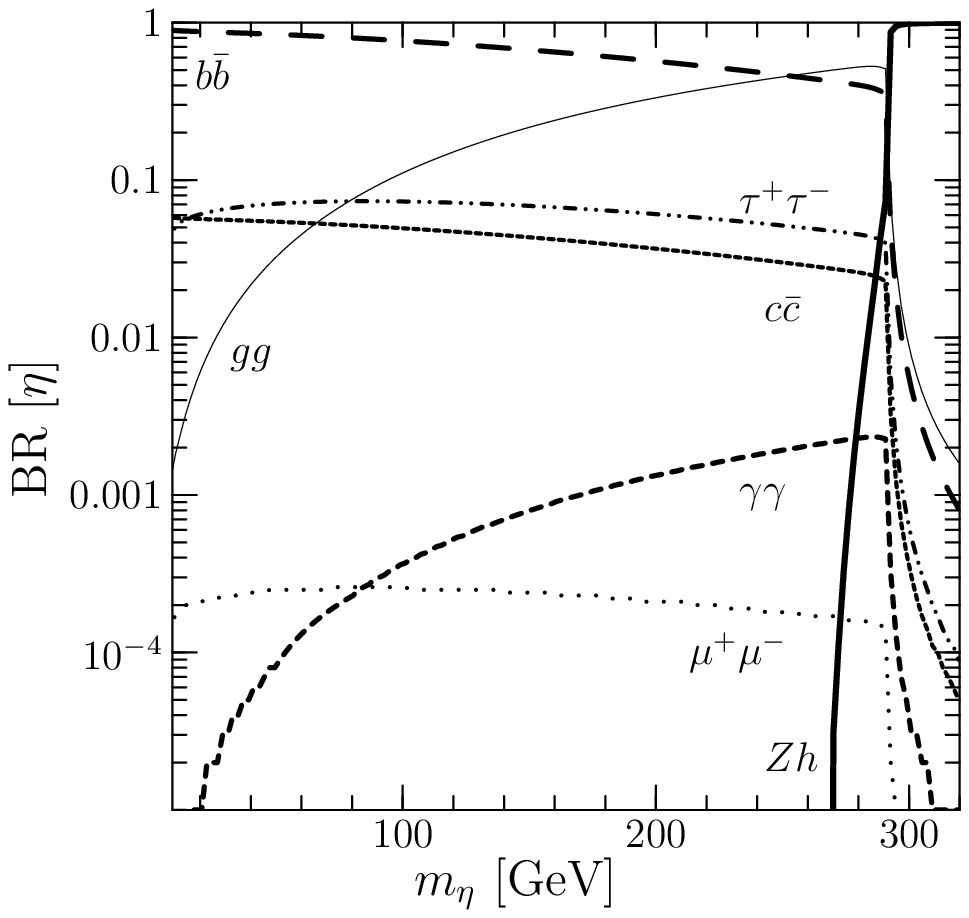}
\end{tabular}
\end{center}
\vspace{-5mm}
\caption{Typical values for the pseudo-axion and Higgs masses in the
Simplest Little Higgs model as a function of the explicit symmetry
breaking parameter $\mu$.  Upper pair: left, Higgs mass for fixed
$F_2=2$~TeV; right, for fixed $\tan\beta=4$.  Lower pair: left, $\eta$
mass; right, $\eta$ branching fractions.}
\label{fig:par_space}
\end{figure}

We begin with the so called ``Golden Point'' discussed in
Ref.~\cite{simplest}, where
\bq
F_1 = 0.5 \,\TeV, \quad F_2 = 2 \,\TeV \quad (\tan\beta=4),
\quad \Lambda = 5 \,\TeV
\eq
which yields $m_T=1000$~GeV (relevant for calculating the $gg\to\eta$
rate).  The Higgs and $\eta$ masses as a function of $\mu$ are shown
by the $\tan\beta=4$ curves on the upper pair of and lower left panels
of Fig.~\ref{fig:par_space}, respectively.  In the lower right panel
of the same figure are the $\eta$ branching ratios.  For large masses,
the decay to $ZH$ utterly dominates, with a very sharp transition to
this regime.  For most other mass values the $b\bar{b}$ decay
dominates, although there is a corridor where $gg$ is large, for
$\eta$ masses between $200$ and $250$~GeV.  Also, compared to the
Higgs, $\gamma\gamma$ is sizable over almost all of the parameter
range.  This is due to the enhanced loop factors from the
pseudo-axion--top quark couplings.  For more details, cf.~Sec.~7.4. of
Ref.~\cite{CPNSH}, and Ref.~\cite{pseudoaxion}.

\medskip

There are three scenarios to consider at LHC:
\begin{itemize}
\item[1.] $m_H>m_\eta+m_Z$: resonant heavy Higgs production with decay
to on-shell $\eta$ and $Z$.  The only likely channel which could be
reliably seen above QCD backgrounds and still yield reconstruction of
the Higgs boson would be $\eta\to b\bar{b}$ and $Z\to\ell^+\ell^-$.
Leptonic $Z$ decay would be necessary for triggering and clean
identification of the $Z$ resonance.  The final state is then
$b\bar{b}\ell^+\ell^-$ ($\ell=e,\mu$).
\item[2.] $m_\eta>m_H+m_Z$, $m_H\lesssim 145$~GeV: resonant pseudo-axion
production with decay to on-shell Higgs and $Z$.  A light Higgs boson
will decay primarily to $b\bar{b}$, necessitating once again leptonic
$Z$ decay.  Thus, the final state signature is the same as case \#1,
and the same analysis will apply.
\item[3.] $m_\eta>m_H+m_Z$, $m_H\gtrsim 145$~GeV: as in case \#2, but
the Higgs preferentially decays to a $W$ boson pair, which further
decay to SM fermions.  The dual-hadronic decay channel is overwhelmed
by QCD background ($Z$+jets), but the branching ratio for one leptonic
and one hadronic decay is almost as large as the dual-hadronic mode,
but suffers from very little QCD background.  The final state is then
$jj\ell^+\ell^-\ell^\pm$, with a same-flavor opposite-sign pair
reconstructing to the $Z$ mass.
\end{itemize}
We perform analyses to cover the two dominant final state scenarios,
considering only resonant production.  While there is significant
parameter space where neither the $\eta$ nor the Higgs boson can be
resonant, the large-QCD background environment of LHC would prohibit
its observation.  If Little Higgs is realized in nature and parameter
space lies in this regime, only a future ILC would be able to detect
this coupling.  We study this in Sec.~\ref{sec:ILC}.


\subsection{The $b\bar{b}\ell^+\ell^-$ final state analysis}
\label{sub:bbll}

This analysis covers both scenarios \#1 and \#2 above, where the $Z$
boson decays to an $e$ or $\mu$ pair and the daughter Higgs boson or
pseudo-axion decays to $b\bar{b}$.  If the Higgs boson is the
daughter, it must be quite light so will likely require many tens of
inverse femtobarn integrated luminosity (several years running) to
discover in a standard production channel~\cite{Cranmer:2005mc}.  This
is a fairly clean signature, as charged leptons are identifiable with
extremely high efficiency and are well-measured, while the $b$ quarks
give displaced vertices which are taggable by the detector, largely
separating them from general QCD jet backgrounds.  Except for
typically small missing transverse momentum from the $b$ decays, the
final state is fully reconstructible, yielding a lepton pair and the
very narrow $Z$ boson peak (no dilepton continuum), a Higgs boson or
pseudo-axion daughter resonance in $b$ jet pairs, and a resonance in
the $b\bar{b}\ell^+\ell^-$ invariant mass.

We must consider backgrounds from any process which produces the same
final state.  This is dominantly continuum QCD $b\bar{b}\ell^+\ell^-$
production, and the small fraction of $t\bar{t}$ production which has
very little missing momentum.  We calculate both of these using matrix
elements generated by {\sc madgraph}~\cite{Stelzer:1994ta}, which
includes all spin correlations through production and decay, although
both inclusive rates are known to at next-to-leading order at
LHC~\cite{Campbell:2003hd,tt_NLO+} and we will later apply K-factors
to account for the corresponding rate enhancements.

To satisfy the basic detector requirements of observability and
trigger, we impose the following kinematic cuts~\cite{ATLAS,CMS}:
\ba\label{eq:cuts-bbll-1}
p_T(b)    > 25~{\rm GeV} \; , & |\eta(b)|    < 2.5 \\\notag
p_T(\ell) > 15~{\rm GeV} \; , & |\eta(\ell)| < 2.5 \\\notag
\triangle R(bb,b\ell) > 0.4 \; , & \triangle R(\ell\ell) > 0.2 \; .
\ea
To identify the $Z$ peak we furthermore require that
\bq\label{eq:cuts-bbll-2}
89.6 < m_{\ell\ell} < 92.8~{\rm GeV} \; ,
\eq
which corresponds to $68\%$ capture of an on-shell $Z$ decay.  To
reduce the top quark background we require very little observed
missing transverse momentum,
\bq\label{eq:cuts-bbll-3}
\sla{p}_T < 30~{\rm GeV} \; .
\eq
To roughly simulate detector effects, we apply Gaussian smearing of
the $b$ jets and charged leptons according to CMS
expectations~\cite{CMS}, as well as missing energy in $b$ jet decays
according to a known distribution.  This does not replace a full
detector simulation, but does make our estimates more realistic,
especially in terms of smeared-out invariant masses, to which we will
apply a fixed window to isolate the $b\bar{b}\ell^+\ell^-$ resonance
peak.

\begin{figure}[ht!]
\begin{center}
\begin{tabular}{lr}
\includegraphics[scale=0.57]{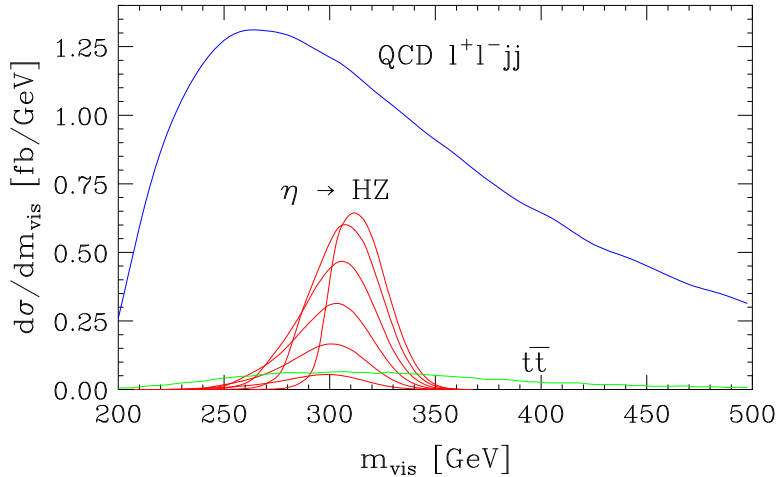}
\end{tabular}
\end{center}
\vspace{-6mm}
\caption{Total invariant mass distribution at the LHC of the 
$b\bar{b}\ell^+\ell^-$ final state for the Little Higgs signal (red),
QCD continuum (blue) and top quark pair (green) backgrounds, using the
cuts of
Eqs.~(\protect\ref{eq:cuts-bbll-1}-\protect\ref{eq:cuts-bbll-3}).
Signal results are for the Golden Point with $\mu$ varying between 147
and 152~GeV.}
\label{fig:m_vis.base}
\end{figure}

Together these cuts result in a QCD $b\bar{b}\ell^+\ell^-$ background
of 297~fb and a top quark background of 10.8~fb.  Clearly, the QCD
continuum dominates, as expected, due to the stringent missing
transverse momentum restriction.  The signal cross section is affected
by the cuts, typically with something like a 1/4--1/3 loss, but this
varies depending on parameter choices (Higgs and pseudo-axion masses)
and is not necessary to detail.  As an example, for the Golden Point
with $\mu=150$~GeV, which results in $m_H=132$~GeV and
$m_\eta=309$~GeV, the cross section with cuts is 20.5~fb.  For all our
results we count $Z$ boson decays to both $e^+e^-$ and $\mu^+\mu^-$.
To illustrate the resonance feature we show in
Fig.~\ref{fig:m_vis.base} the differential cross sections (without ID
efficiencies) with respect to the visible invariant mass for signal
and background, using the Golden Point parameterization and with
$147<\mu<152$~GeV.  For these choices the pseudo-axion is the parent.
Recall from Fig.~\ref{fig:par_space} that $m_\eta\propto\mu$, while
the Higgs mass is inversely proportional, but in a more complicated
way that includes a plateau region in mass at low $\mu$.  For
$\mu=147$~GeV, the Higgs mass is 158~GeV, where there is almost no
branching ratio (BR) to $b\bar{b}$, resulting in a very small rate.
This is clearly in the region where one should perform instead an
$jj\ell^+\ell^-\ell^\pm$ analysis, looking for the $H\to W^+W^-$
decay; we address this channel in the next subsection.

\begin{figure}[hb!]
\begin{center}
\includegraphics[scale=0.9]{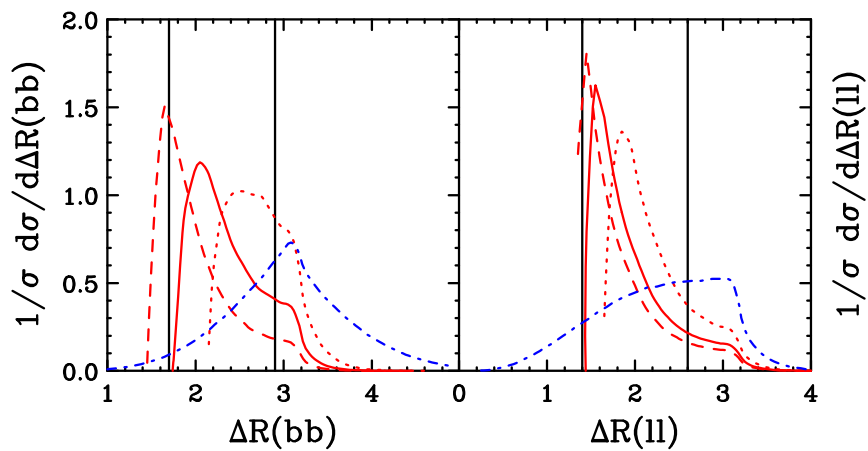}
\end{center}
\vspace{-6mm}
\caption{Normalized angular correlations in the lego plot separation 
for the $b$ jet pair (left) and lepton pair (right).  The signal is
shown in red for the Golden point and $\mu=147,150,152$~GeV using
dotted, solid and dashed lines, respectively.  The QCD continuum
backgrounds is shown in blue (dot-dashed).  Vertical lines represent
cuts used for the $\mu=150$ case as described in the text.}
\label{fig:bbll-ang}
\end{figure}

It is possible to perform a simple ``bump-hunting'' analysis, although
it is far from optimum.  We observe significant angular correlations
in the leptons and $b$ jets which can be used to reject the
backgrounds further.  Fig.~\ref{fig:bbll-ang} shows the two most
important correlations, the $b-b$ and $\ell-\ell$ lego plot
separations.  There are distinct differences between resonant and
continuum production.  For the Golden Point and $\mu=150$~GeV, it is
possible to reduce the QCD background by a factor 3 while losing only
$15\%$ of the signal.  This does sculpt the total invariant mass
distribution somewhat, but reduces by about half the amount of
luminosity required for discovery.  In general, though, the angular
cut choice needed changes with choice of input parameters, as it
depends ultimately on the decay kinematics -- how boosted the
daughters are.  We don't attempt such a complicated analysis here, but
point out the correlations for future detailed work at the detector
level.
\begin{figure}[ht!]
\begin{center}
\includegraphics[scale=0.75]{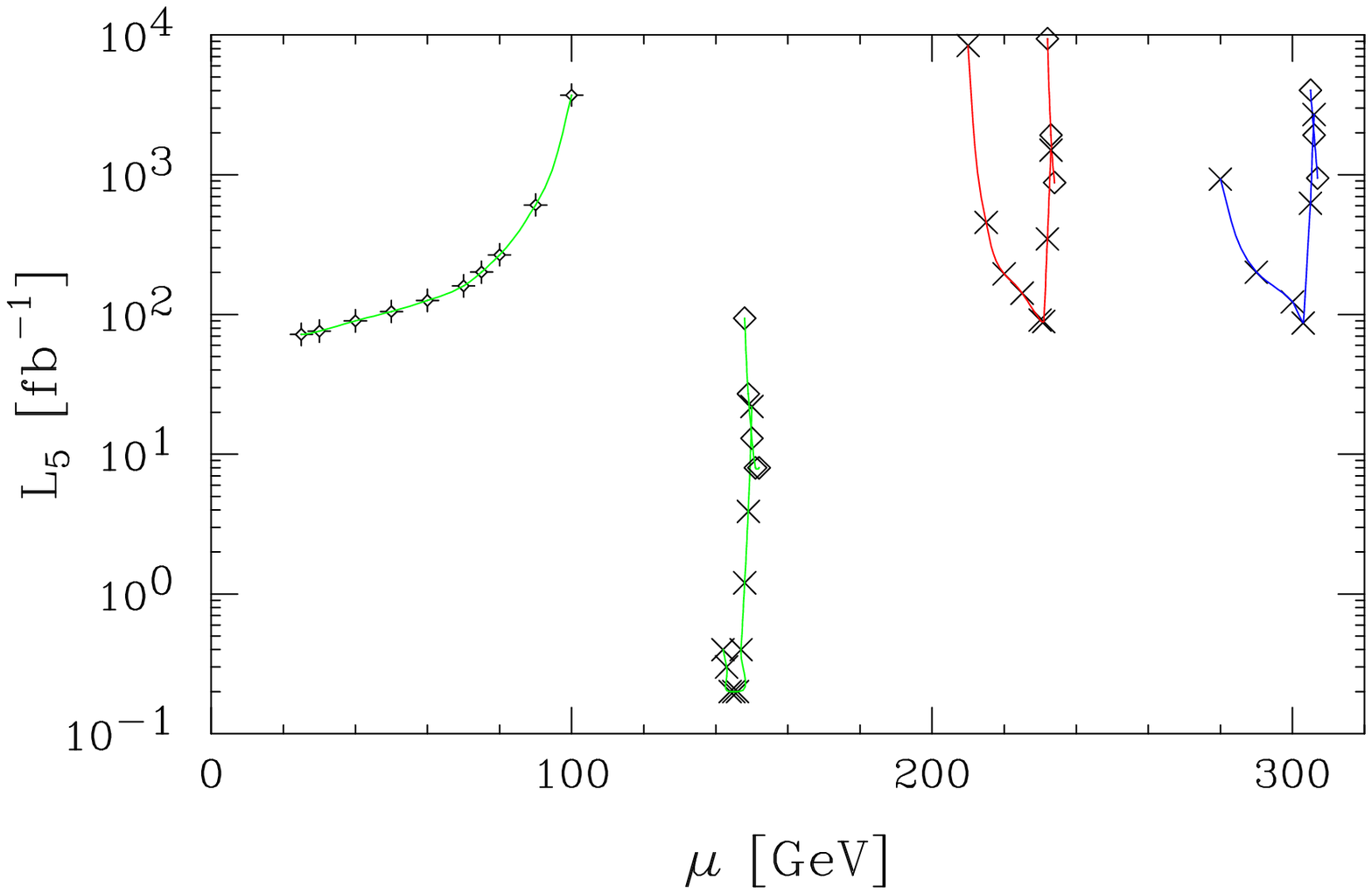}
\end{center}
\vspace{-6mm}
\caption{Required integrated luminosity for one LHC detector to make
a $5\sigma$ observation of the Little Higgs pseudo-axion.  The Golden
Point is green, while blue (red) are for $F_{1,2}=1,2\,(1,4)$~TeV.
Crosses represent the $b\bar{b}\ell^+\ell^-$ analysis for $gg\to
H\to\eta Z$, while diamonds are the same for $gg\to\eta\to HZ$
(cf. Subsec.~\protect\ref{sub:bbll}), and X symbols are for
$gg\to\eta\to HZ$ where the Higgs boson is heavier and decays to
$jj\ell\nu$ (cf. Subsec.~\protect\ref{sub:3l2j}).  Crosses for the
larger $F_{1,2}$ choices are above $10^4$.}
\label{fig:LHC-lumi}
\end{figure}

We present our overall results then, without optimization using
angular correlations, in Fig.~\ref{fig:LHC-lumi} in terms of the
required luminosity for one LHC detector to make a $5\sigma$
observation of the $b\bar{b}\ell^+\ell^-$ final state above QCD
backgrounds.  To obtain this we used ID efficiency factors of $50\%$
for each $b$ jet and $95\%$ for each charged lepton, and a capture
efficiency of $68\%$ for the dilepton mass window (signal only).  We
also include K-factors for signal and background to take into account
the large QCD rate corrections: 2.3 for the signal~\cite{H-NNLO}, 1.3
for the QCD $Zb\bar{b}$~\cite{Campbell:2003hd} background, and 1.4 for
$t\bar{t}$~\cite{tt_NLO+}.  Our results are only estimates, not the
optimal reach, and not comprehensive over parameter space.  In
addition to the Golden Point, we show two other cases for different
choices of $F_{1,2}$.  For $F_{1,2}=1.0,2.0$~TeV ($\tan\beta=2$), the
collective scale $F^2=F_1^2+F_2^2$ is approximately the same, but the
$Z$--$H$--$\eta$ coupling squared is reduced by an order of magnitude.
For $F_{1,2}=1.0,4.0$~TeV ($\tan\beta=4$, same as the Golden Point),
the scale $F$ is about twice as large, but the coupling strength is
unchanged.

For $H\to Z\eta$, shown by crosses, large $F$ values drive $m_H$ to
several hundred GeV.  The $Z\eta$ BR drops rapidly with increasing
Higgs mass, as it is out-competed by the $VV$ modes' double
longitudinal-polarization enhancement.  Thus a heavy Higgs boson
decaying to $Z\eta$ would likely be observable with LHC luminosity
only for fairly small values of $F_{1,2}$ and $\mu$.

The $\eta\to ZH$ case suffers a somewhat different fate, as shown by
the diamonds.  Typically for any given $F_{1,2}$ there is only a very
small range of $\mu$ which can be addressed.  If $\mu$ is too large,
$m_H$ is smaller than the LEP limit.  Only a few GeV lower than this
limit, the Higgs boson mass lies in the region where decays to weak
bosons dominate, thus the BR to $b\bar{b}$ drops to zero and the
$jjjj\ell^+\ell^-$ analysis takes over.  For larger $F_{1,2}$ than at
the Golden Point, $m_\eta$ increases, lowering the production rate,
even though the BR to $ZH$ is still large.


\subsection{The $jj\ell^+\ell^-\ell^\pm$ final state analysis}
\label{sub:3l2j}

For $m_H\gtrsim 140$~GeV, the BR to $b\bar{b}$ drops off dramatically,
being replaced mostly by $W^+W^-$.  This requires a different
analysis.  Because the $W$ boson leptonic BR is smaller than that to
quarks (jets), and we already have trigger leptons in the final state
from the very sharp $Z$ boson resonance, the obvious final state to
consider is the largest in terms of BR: $W^+W^-\to jjjj$.  It is fully
reconstructible, and typically better so than the
$b\bar{b}\ell^+\ell^-$ case, because light-flavor jets typically give
much less missing energy.  However, this channel suffers from a QCD
$Z$+jets background of several hundred fb~\cite{Mangano:2002ea}.  A
quick analysis of this channel shows that in principle it would be
statistically possible for some parameter space, but the pseudo-axion
resonance would peak at about the same places the QCD continuum does,
with a $S/B$ ratio of about 1/50, which makes prospects dodgy.

Instead, we investigate the $H\to WW\to\ell\nu jj$ channel, which has
2/3 the BR of the all-hadronic channel, but about 1/100 the
background, which comes almost exclusively from $ZWjj$
production~\cite{Barger:1989yd}.  The final state is then three
charged leptons, two jets and missing transverse momentum.  The basic
kinematics cuts for detector acceptance and trigger are:
\ba\label{eq:cuts-4jll-1}
p_T(j)    > 20~{\rm GeV} \; , & |\eta(j)|    < 4.5 \; , \\\notag
p_T(\ell) > 15~{\rm GeV} \; , & |\eta(\ell)| < 2.5 \; , \\\notag
\triangle R(jj,j\ell) > 0.4 \; , & \triangle R(\ell\ell) > 0.2 \; , \\\notag
89.6 < m_{\ell\ell} < 92.8~{\rm GeV} \; , & 
\sla{p}_T < 30~{\rm GeV} \; .
\ea
The $Z$-pole cut is the same as before, applied on the $Z$ decay
products explicitly (combinatorics will be a very minor correction).
Typically only about $25\%$ of the signal survives, but the total
continuum background is now only a factor of two larger than the
signal for the Golden Point and $\mu=145$~GeV.

\begin{figure}[ht!]
\begin{center}
\includegraphics[scale=0.57]{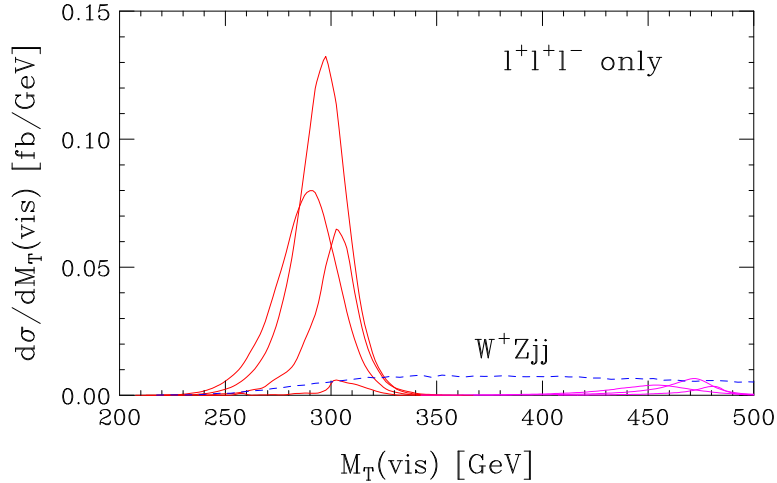}
\end{center}
\vspace{-6mm}
\caption{Transverse mass distribution of the visible 
$\ell^+\ell^+\ell^- jj\sla{p}_T$ system in $gg\to\eta\to ZH,H\to WW$
resonant production at the LHC.  Red and magenta curves are for the
Golden Point and $F_{1,2}=1.0,2.0$~TeV, respectively, and the dashed
blue curve is the $W^+Zjj$ background.}
\label{fig:mT}
\end{figure}

With only one neutrino, it is straightforward to construct a
transverse mass for the entire system, which peaks very close to the
pseudo-axion mass as expected, with only minor smearing due to
detector effects.  We show examples for illustration in
Fig.~\ref{fig:mT} for the Golden Point and $F_{1,2}=1.0,2.0$~TeV, for
a few values of $\mu$.  Since $m_\eta\propto\mu$, larger peak values
of $m_T$ are for larger values of $\mu$.  The Golden Point would
produce a signal far above the background, trivially observable, at
least in the region of $142<\mu<150$~GeV where the $\eta$ can be
resonant and the Higgs boson has at least a modest BR to $WW$.  The
$F_{1,2}=1.0,2.0$~TeV cases yield a signal of the same size as the
background, also quite easy to observe but naturally requiring more
statistics.

This analysis is generally powerful wherever there is a decent signal
rate, due to the small background.  Using the same ID efficiencies and
signal K-factor as in Sec.~\ref{sub:bbll} and a generic QCD K-factor
of 1.3 for the background ($WZjj$ is not known at NLO), we summarize
our results again in Fig.~\ref{fig:LHC-lumi} with points using an X
symbol.  The left edge of each curve is cut off by values of $\mu$ for
which the $\eta$ cannot be resonant, so the rate is hopelessly small.
As $\mu$ increases, $m_\eta$ increases and $m_H$ decreases, opening up
more phase space for the decay, resulting in more events passing the
cuts, which in turn means less luminosity required.  However, below
$H\to WW$ threshold the BR to $WW$ begins to falls off steeply,
resulting instead in an increase in required luminosity with
increasing $\mu$.  At some point, the $H\to b\bar{b}$ analysis becomes
more powerful, and the curves cross those with the diamond points.

%


\section{Phenomenology at an ILC}
\label{sec:ILC}

\begin{figure}[ht!]
\begin{center}
\includegraphics[scale=.8]{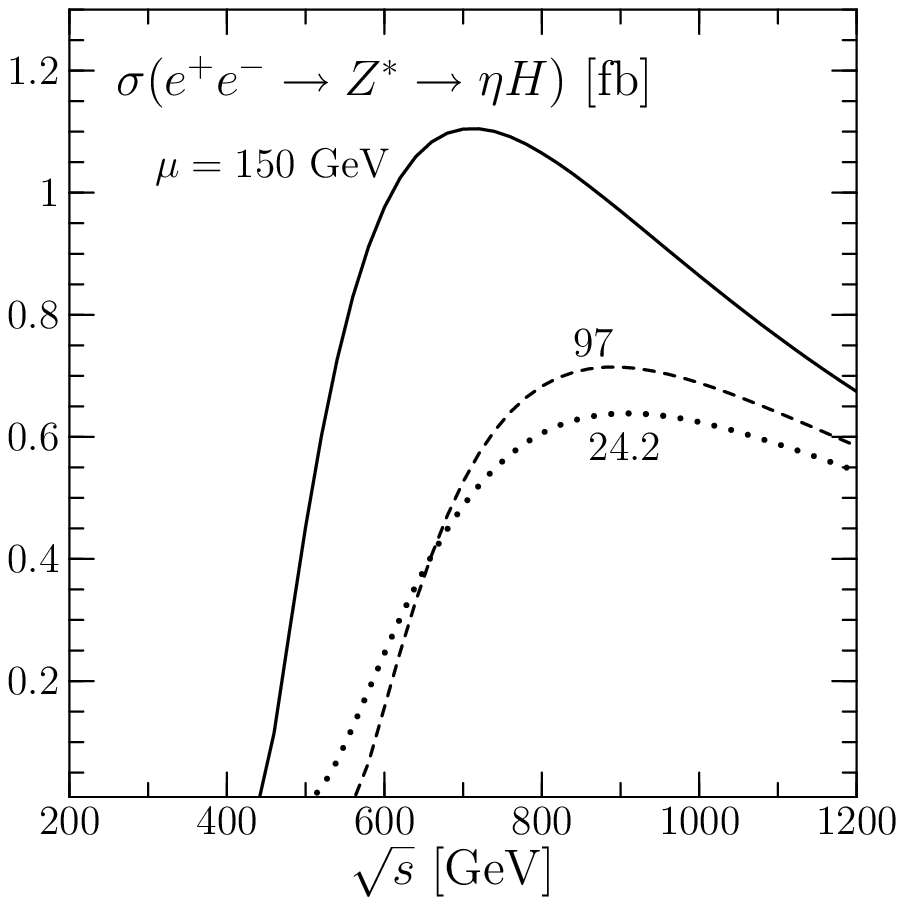} \quad
\includegraphics[scale=.8]{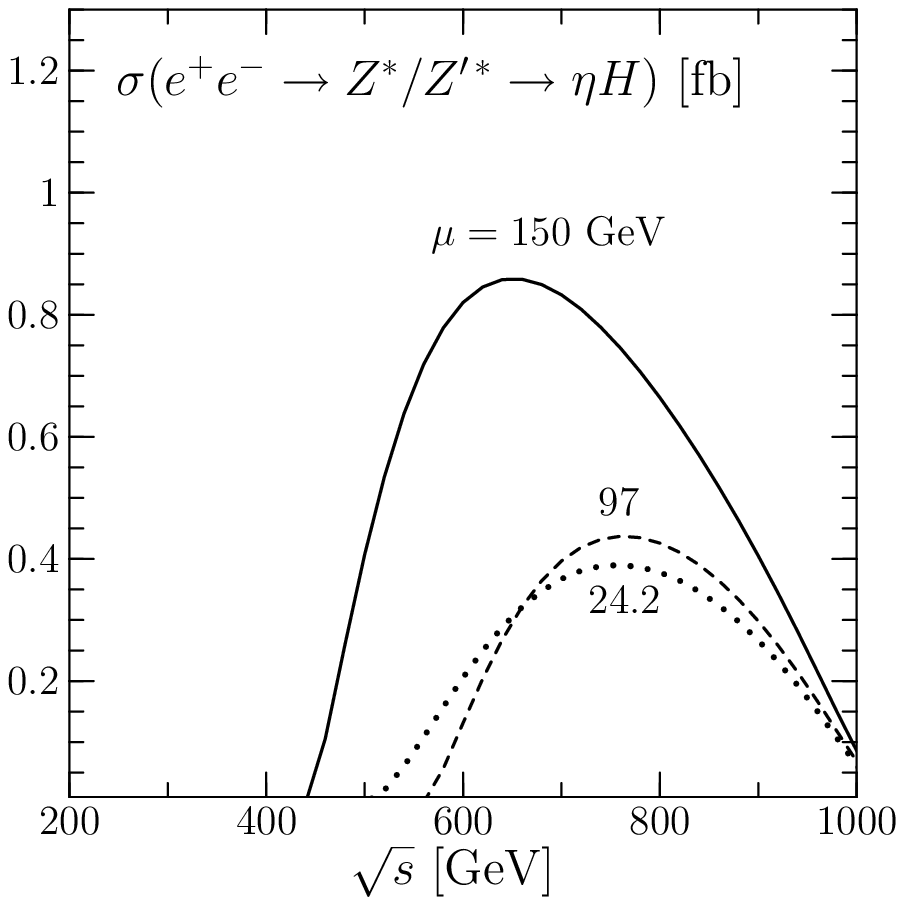}
\end{center}
\vspace{-5mm}
\caption{Total cross section for $H\eta$ production at an ILC for
different values of $\mu, m_\eta, m_h$: the Golden Point with
$150/309.2/131.7$~GeV (solid), $97/200.0/368.4$~GeV (dashed), and
$24.2/50.0/451.3$~GeV (dotted).  The left panel is without the
presence of a $Z'$ resonance, while the right panel includes
destructive $Z$--$Z'$ interference for $m_{Z'} = 1152$~GeV.}
\label{fig:ilc_totcross}
\end{figure}

At a future ILC, one would not have to rely on on-shell production to
prove or place a limit on the existence of a $Z$--$H$--$\eta$
coupling.  As long as it is present and $H\eta$ pair production is
kinematically allowed, the pseudo-axion could be seen.
Fig.~\ref{fig:ilc_totcross} shows the total cross section at an ILC
for various $\sqrt{s}$ for this channel at the Golden Point for three
different values of $\mu=24.2/97/150$~GeV, for which
$m_\eta=309.2/200/50$~GeV and $m_H=131.7/368.4/451.3$~GeV,
respectively.  The maximum cross section is of the order of
0.4--1.2~fb~$\times\tan^2\beta$, as shown in the left panel of
Fig.~\ref{fig:ilc_totcross}.  Since in the Simplest Little Higgs there
is a destructive interference between the SM $Z$ boson and the $Z'$
boson, the maximum total cross section in the full model goes down by
roughly a quarter, as shown in the right panel.

\begin{figure}
\begin{center}
\includegraphics[scale=.8]{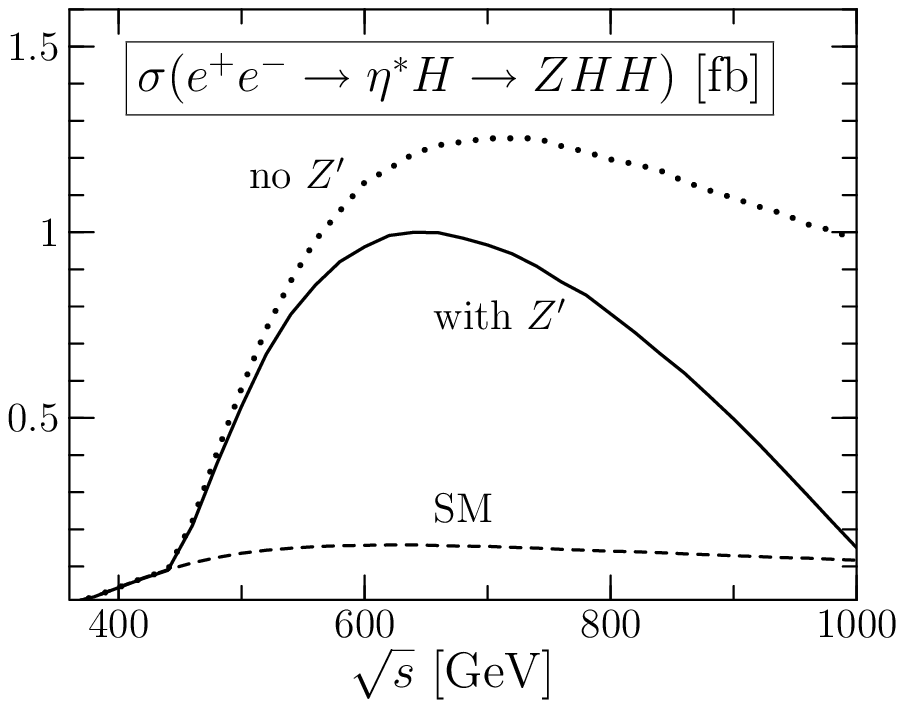} \quad
\includegraphics[scale=.8]{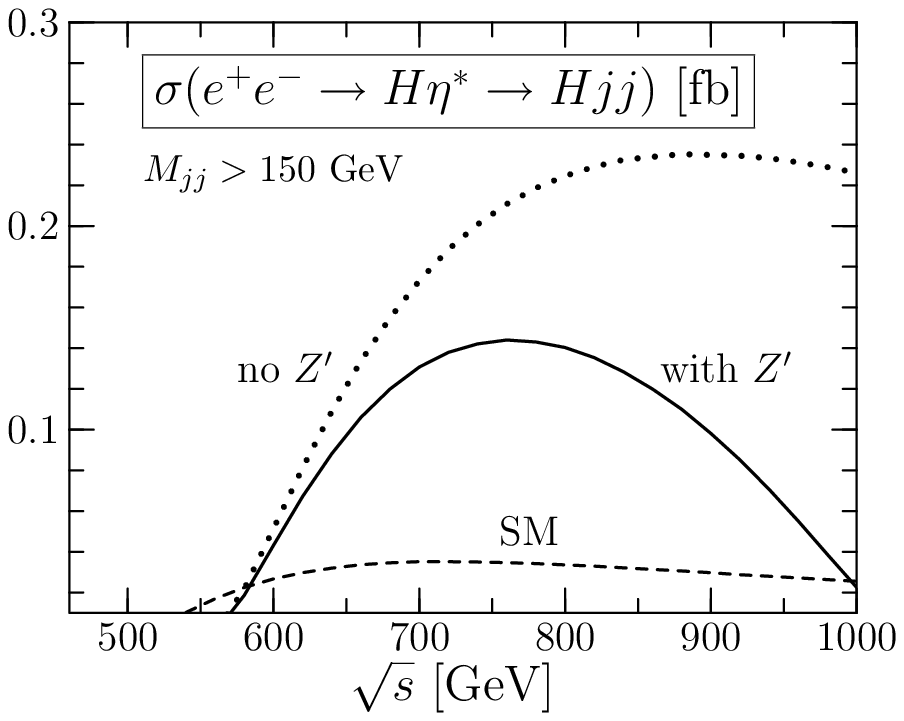}
\end{center}
\vspace{-5mm}
\caption{Cross section in the $\mu$ model pseudo-axion with (solid
line) and without (dotted line) the $Z'$ resonance at the Golden
Point.  The dashed line is the SM prediction.  On the left is the
$ZHH$ final state for $\mu=150$~GeV, while the right is the $Hjj$
final state for $\mu=97$~GeV.}
\label{fig:finalzhh}
\end{figure}
\begin{figure}
\begin{center}
\includegraphics[scale=.8]{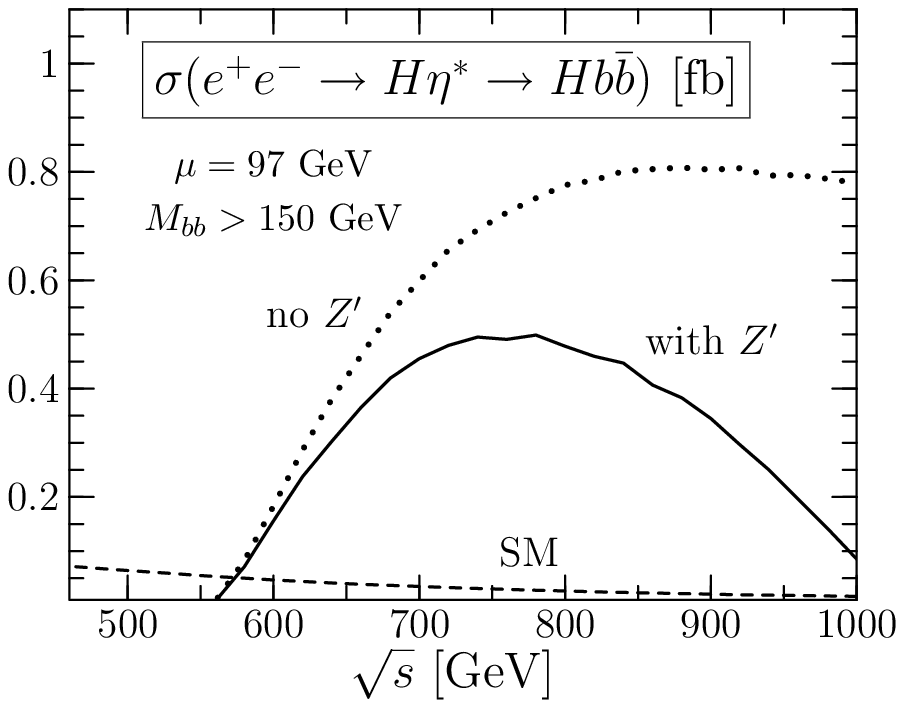} \quad
\includegraphics[scale=.8]{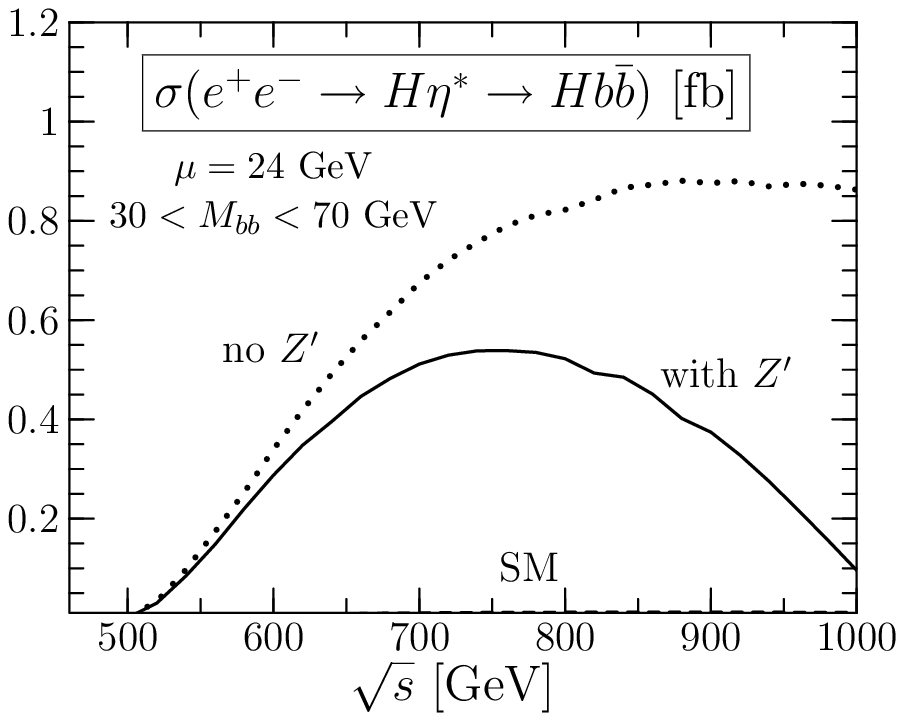}
\end{center}
\vspace{-5mm}
\caption{Cross section for $Hb\bar{b}$ production in the $\mu$--model 
pseudo-axion with (solid line) and without (dotted line) the $Z'$
resonance at the Golden Point.  The dashed line is the SM prediction.
On the left is $\mu=97$~GeV, right $\mu=24.2$~GeV.}
\label{fig:finalhbb}
\end{figure}

Depending on the mass of the pseudo-axion, the main decay channel for
the $\eta$ varies from (cf. Fig.~\ref{fig:par_space}) $b\bar{b}$ when
fairly light, to two gluons (jets) for intermediate masses, and
finally only $ZH$ for masses above that threshold, about
$m_\eta\gtrsim 290$~GeV at the Golden Point.  For such heavy
pseudo-axions, one would look into the $ZHH$ final state.  The cross
section for this process is shown in the left panel of
Fig.~\ref{fig:finalzhh} (this and the other ILC simulations were
performed by the BSM extensions of the event generator {\sc
O'Mega/Whizard}~\cite{Omega,Whizard}).  The dashed curve is the SM
cross section, the dotted line shows the signal process without a $Z'$
resonance or with one above 1.5~TeV, while the solid line is the
prediction for $M_{Z'}=1.15$~TeV.  For the parameter values considered
here the Higgs mass is $131.7$~GeV, so that the six-fermion final
state of interest consists of four $b$ jets and a lepton pair.  The SM
$e^+e^-\to ZHH$ process was extensively studied to analyze the triple
Higgs coupling measurability at an ILC~\cite{TDR,Djouadi:1999gv}.  It
was found that the search limit for this process is severely
statistics-limited and hence hard to analyze.  Fig.~\ref{fig:finalzhh}
shows that even for the presence of a destructive interference with a
$Z'$ boson, the cross section in the Simplest Little Higgs is larger
than in the SM by a factor of five, allowing for a reliable analysis
of the final state.  The pseudo-axion shows up as a sharp spike in the
$bb\ell\ell$ spectrum, since even if the $ZH$ decay is kinematically
allowed the width is of the order of a few hundred MeV and hence far
smaller than any conceivable detector resolution.

If the pseudo-axion mass lies between roughly $200$ and $290$~GeV, the
$\eta$ has a large branching fraction into a gluon pair, of the order
of $30-50\%$, which can even become dominant close to the $ZH$
threshold.  For this parameter range, the Higgs boson decays
completely to $WW$ or $ZZ$, so the complete final states are
$jj\ell\ell\ell\ell$, $jjjj\ell\ell$ or six jets.  Higgs bosons in an
intermediate mass range could be reconstructed at an ILC~\cite{TDR}
fairly well, and although there would be combinatorial background in
the $H\to 4j$ channel, this would not degrade the pseudo-axion search.
The right panel of Fig.~\ref{fig:finalzhh} shows the cross section to
$Hjj$, where a cut on the dijet invariant mass has been applied to
reduce the SM Higgsstrahlung background which is shown as a dashed
line.  Again, the dotted line is the prediction without a low-lying
$Z'$, while the solid line shows the destructive $Z$--$Z'$
interference.  For the Golden Point with $F_2=2$~TeV and
$\tan\beta=4$, there is a critical point at $\mu\sim 128$~GeV where
the pseudo-axion and Higgs boson are degenerate (at $\sim 264$~GeV).
But vector boson-mediated production cannot yield two identical
particles (under the assumption of CP conservation); hence this must
be either a scalar and a pseudoscalar or a scalar and a vector.  From
the normalization of the $b\bar{b}$ peak, which does not fit to the
Higgs boson, one could deduce that on one side of the event there must
be a degenerate pseudoscalar present.

The best case is where $b\bar{b}$ is the dominant $\eta$ decay mode,
i.e. for pseudoscalar masses below 250~GeV.  The SM background is
mainly Higgsstrahlung with $Z\to b\bar{b}$ decay, which can be
distinguished from the signal by applying a cut to the $b\bar{b}$
invariant mass.  For the parameter space under consideration, the
Higgs decays to a vector boson pair which can be measured
semi-leptonically.  Fig.~\ref{fig:finalhbb} shows the cross section
for the production of a Higgs boson in association with a $b\bar{b}$
pair, on the left for a pseudo-axion mass of 200~GeV, on the right for
an extremely light $\eta$ with $m_\eta=50$~GeV.  For the former case
one can cut out all bottom quark pairs below 150~GeV to get rid of the
$Z$ background; while this is not a very rigorous cut, it already
reduces the SM background to a negligible level.  To extract the light
pseudoscalar we adopt a window between 30 and 70~GeV for $m_{bb}$.
Again, Fig.~\ref{fig:finalhbb} shows as a dotted line the cross
section without $Z'$ interference, and solid with the
$m_{Z'}=1.15$~TeV.  In both cases a few hundred $\fb^{-1}$ should
suffice to observe $H$--$\eta$ production in that channel and thus
establish the presence of a $Z$--$H$--$\eta$ coupling.


\section{Conclusions}

We have identified an important distinction between Product Group and
Simple Group Little Higgs models crucially involving the pseudo-axion
corresponding to ungauged diagonal generators in the former case, and
the pseudoscalar combination of anomalous $U(1)$ diagonal generators
in the latter.  By consideration of quantum numbers alone, a coupling
between the pseudo-axion, Higgs boson and $Z$ gauge boson is allowed.
However, by construction such a coupling must vanish to all orders in
Product (Gauge) Group models, where the {\em global} Little Higgs
symmetry group is simple.  In Simple Group models it is proportional
to the difference between the vevs of the chiral field which occur in
pairs.  Observing resonant production at the LHC, either $\eta\to ZH$
or $H\to Z\eta$, or Drell-Yan $H\eta$ production in $e^+e^-$
collisions, would conclusively rule out a Product Group realization of
the Little Higgs mechanism were a candidate Little Higgs found at the
LHC.  Note, however, that this coupling would also vanish in Simple
Group models with an imposed T-parity~\cite{Martin:2006ss}, as the
discrete symmetry forces $F_1=F_2$.

The phenomenology at LHC is in principle straightforward, breaking
down into three classes, depending on whether the pseudo-axion or
Higgs boson is heavier, and whether the Higgs boson decays primarily
to a pair of bottom quarks or $W$ gauge bosons.  We find that
observation of $gg\to\eta\to ZH\to\ell^+\ell^-b\bar{b}$ is in general
possible only for a very small range in $\mu$, where the Higgs boson
is light enough to decay to bottom quarks but is not yet ruled out by
the LEP results.  For small values of $F$, 1--2~TeV, this could be
achieved at LHC with a reasonable few hundred inverse femtobarn of
luminosity, but for large values of $F$, a few TeV, the $\eta$ is more
massive thus harder to produce, so only the luminosity-upgraded LHC
(SLHC) would have a chance of observation of this coupling.  (This
would furthermore probably require different cuts, such as a less
restrictive $\sla{p}_T$ cut in the $b\bar{b}\ell^+\ell^-$ analysis,
which likely would allow much more background from $t\bar{t}$.)  For
similar values of $F_1$ and $F_2$, the coupling is too small.  If the
Higgs boson is heavier, one generally loses as it preferentially
decays to two longitudinal gauge bosons rather than $Z\eta$, so the BR
drops to negligible values.  If the pseudo-axion is resonant but the
Higgs boson decays to gauge bosons, a multi-lepton analysis works
extremely well, especially for $F$ around 1--2~TeV: even a few inverse
femtobarn cover much of the possible parameter space.  However, we
find that LHC has large gaps in coverage, for regions where neither
the Higgs boson nor the pseudo-axion can be resonant.  This is
precisely where a linear collider would be crucial.

At a future linear collider the detection of the pseudo-axion would be
quite easy using the $Z$--$H$--$\eta$ coupling in scalar pair
production.  The cross section for the whole parameter space is large
enough to give several hundred pseudo-axions for reasonable
luminosity, irrespective of any of $\eta\to ZH$ or $H\to Z\eta$ being
on-shell.  This closes the holes in the LHC discovery range.  An
important point is a destructive $Z/Z'$ interference in that channel
which is most severe for the lowest $F$ scales but does not jeopardize
the signal.  In the case of an on-shell decay $\eta\to ZH$ the $ZHH$
final state is enhanced compared to the SM by a factor of at least two
for the region favored by electroweak precision data and up to an
order of magnitude for large $\tan\beta$.  The same holds for the
$b\bar b$ decay mode of the pseudo-axion where the enhancement over
the SM is less pronounced.  But here the pseudo-axion is easily
visible even for bad signal to background ratio as a sharp peak in the
$bb$ invariant mass spectrum, whose width is completely given by the
detector resolution.  So even for mixing angles as small as
$\tan\beta\sim 1.5$ the pseudo-axion coupling $Z$--$H$--$\eta$
coupling can be detected at an ILC.


\subsection*{Acknowledgments}

We thank Uli Baur for a critical reviews of the
manuscript.  This research was supported in part by the National
Science Foundation under Grant No. PHY99-07949, the U.S. Department of
Energy under grant No. DE-FG02-91ER40685, and by the
Helmholtz-Gemeinschaft under Grant No. VH-NG-005.


\end{document}